\documentclass{elsart}
\usepackage{graphicx,harvard,amssymb}

\def\bibcode#1{(\texttt{#1})}

\def\url#1{{\ttfamily\def\/{/\discretionary{}{}{}}#1}}

\journal{New Astronomy}

\begin{document}
\begin{frontmatter}
\title{On-Axis Orphan Afterglows}
\author[all]{Ehud Nakar\thanksref{en}},
\author[all]{Tsvi Piran\thanksref{tp}}
\thanks[en]{E-mail: udini@phys.huji.ac.il }
\thanks[tp]{E-mail: tsvi@phys.huji.ac.il }
\address[all]{Racah Institute for Physics, The Hebrew
University, Jerusalem, 91904, ISRAEL}

\begin{abstract}
In many GRB inner engine models the  highly  relativistic GRB jets
are engulfed by slower moving matter. This could result in
different beaming for  the prompt $\gamma$-ray emission and for
the lower energy afterglow. In this case we should expect that
some observer will see {\it on-axis orphan afterglows}: X-ray,
optical and radio afterglows within the initial relativistic
ejecta with no preceding GRB; The  prompt $\gamma$-ray emission is
pointing elsewhere. We show that the observations of the WFC on
BeppoSAX constrain with high certainty the prompt X-ray beaming
factor to be less than twice the prompt $\gamma$-ray beaming. The
results of Ariel 5 are consistent with this interpretation.  The
RASS from ROTSE and HEAO-1 constrain the X-ray beaming factor  at
400 and 20 minutes after the burst respectively to be comparable
and certainly not much larger than the $\gamma$-ray beaming
factor. There is no direct limit on the optical beaming. However,
we show that observations of several months with existing
hardware could result in a useful limit on the optical beaming
factor of GRB afterglows.
\end{abstract}
\end{frontmatter}

\section{Introduction}

The realization that Gamma-Ray bursts (GRBs) are beamed changed
our understanding of the phenomenon in many ways. The first and
most dramatic  is the drastic revision of  the bursts' energy.
The enormous $ 10^{54}$ergs, turned out, after beaming
corrections,  to be a ``modest"  $ 10^{51}$ergs. Naturally the
actual GRB rate increased by the inverse factor. Even more
surprising was the discovery (Frail et al., 01; Panaitescu \&
Kumar, 01; Piran et al., 01) of the rather narrow total energy
distribution. Currently, the evidence for beaming is the 'jet
break' in the afterglow's light curve. The jet angle $\theta_j$
is determined from the time of the break in the light curve, that
is interpreted as a jet break (Sari, Piran \& Halpern, 99).
However, even if this interpretation is correct it corresponds to
the jet opening angle when most of the emission is in the optical
or IR bands.  There is no direct evidence for the beaming factor
during the GRB phase, or even during the early afterglow phase
when the emission is mostly in X-rays.  It is not clear whether
the $ \gamma $-ray, X-ray, optical and radio emissions have
similar initial (before the jet spreading) beaming factors. There
are  good physical reasons to question whether there is a common
beaming factor in different wavelengths. Different parts of the
spectrum dominate the emission at different times and correspond
to different physical conditions within the relativistic flow. It
is possible, and some will argue even likely, that emission in
different energy bands will have different beaming factors.

\begin{figure}
\label{fig:OnAxisOrph} {\par\centering
\resizebox*{0.9\columnwidth}{0.5\textheight}{\includegraphics{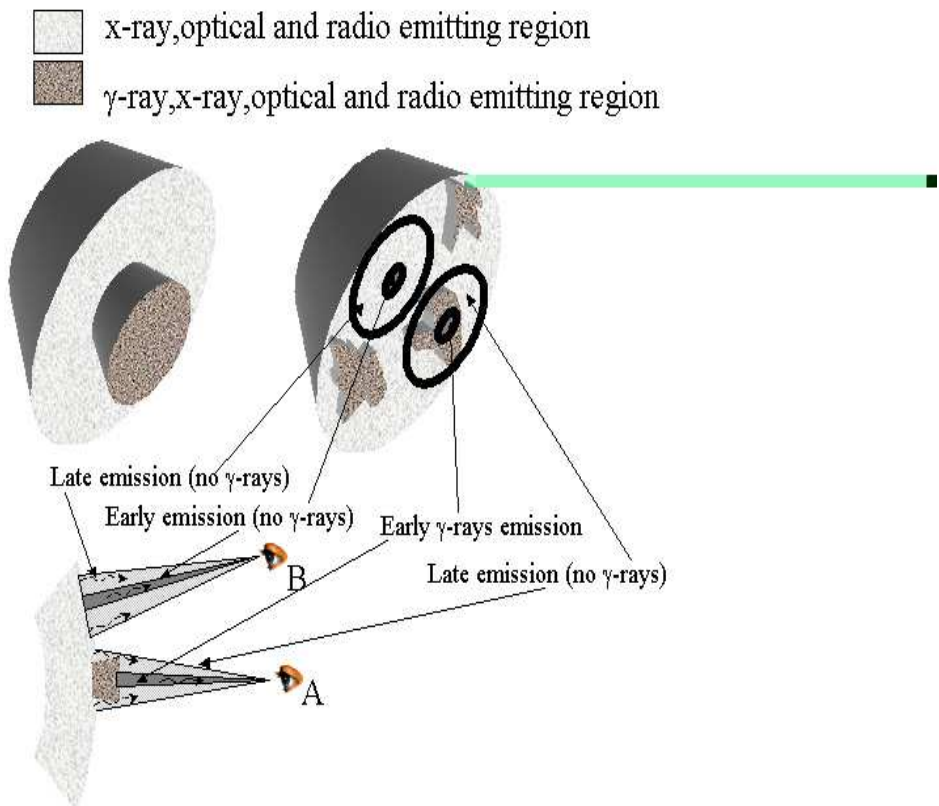}}
\par}
\caption {A schematic description of on-axis orphan afterglow.
Only some regions within the initial ejecta  emit prompt
$\gamma$-ray radiation. The structure of the $\gamma$-ray
emitting regions could be regular (upper left figure) or
irregular  (upper right figure). The ellipses describe the area
observed by an observer at a given time (The smaller ellipses
describe the area observed initially during the $\gamma$-ray
emission, when $\Gamma$ is large, while the larger one describe
the area observed during the X-ray emission, when $\Gamma$ is
smaller.) The lower figure depicts the cross section of the upper
left figure. Observer A detects the early emission from a small
region within the $\gamma$-ray emitting region and later  an
afterglow from a much larger region. This observer would see a
regular GRB and an  afterglow. Observer B does not detect any
$\gamma$-rays but detects a regular afterglow, which we call
on-axis orphan afterglow. }
\end{figure}

The origin of the problem in determining the different beaming
lies in two related relativistic phenomena, casual connection and
relativistic beaming. Regions more than $\Gamma^{-1}$ apart within
the relativistic flow are casually disconnected. Additionally, the
light emitted from a relativistic source moving with a Lorentz
factor $\Gamma$ is beamed into an angle of $\Gamma^{-1}$ along the
line of motion. The first phenomenon implies that regions that
are more than $\Gamma^{-1}$ apart could have different physical
conditions. The second one implies that an observer could see
only one such region at a time and won't know about the other.
Both phenomena imply that the $\gamma$-rays, that are emitted
when $\Gamma \ge 100$, could have come from a jet with an opening
angle of 1/100 and the observer would have no way of telling the
difference from a spherical source. The later X-ray (as well as
optical and radio) afterglow is emitted from a slower moving
material with a lower Lorentz factor. Hence, the X-ray beam could
be larger than the $\gamma$-rays beam. Similarly the region
observed by a given observer is larger as well (see fig
\ref{fig:OnAxisOrph}).

These relativistic effects allow configurations with narrow
$\gamma$-ray beams, wider X-ray and even wider optical and radio
beams. This will produce orphan afterglows - events in which
X-ray, optical and radio afterglow is observed while the prompt
GRB is pointing elsewhere. There are two types of orphan
afterglows: (i) On-axis orphan afterglows (which we introduce and
discussed here, see fig. \ref{fig:OnAxisOrph}) are observed within
the initial relativistic jet by observers that miss the narrower
$\gamma$-ray beam. These afterglows follow  the light curves of
the standard afterglows (observed following regular GRBs). (ii)
Off-axis orphan afterglows (see fig. 2) are the ``traditional"
orphan afterglows (Rohads, 97; Perna \& loeb, 98; Dalal, Griest
\& Pruet 2001, Granot et al. 2002, Nakar, Piran \& Granot 2002,
Totani \& Panaitescu 2002) that are observed outside the initial
jet. Off-axis orphan afterglows can be seen only after the jet
break when the jet expands sideway. Their light curve rises
initially reaching a maximal flux (that depends on the observing
angle) and then decays following the post-jet-break light curves
of  a standard GRB afterglow. To study the initial opening angles
of the relativistic jets we must consider the on-axis orphan
afterglows.

\begin{figure}
\label{Off} {\par\centering
\resizebox*{0.9\columnwidth}{0.4\textheight}{\includegraphics{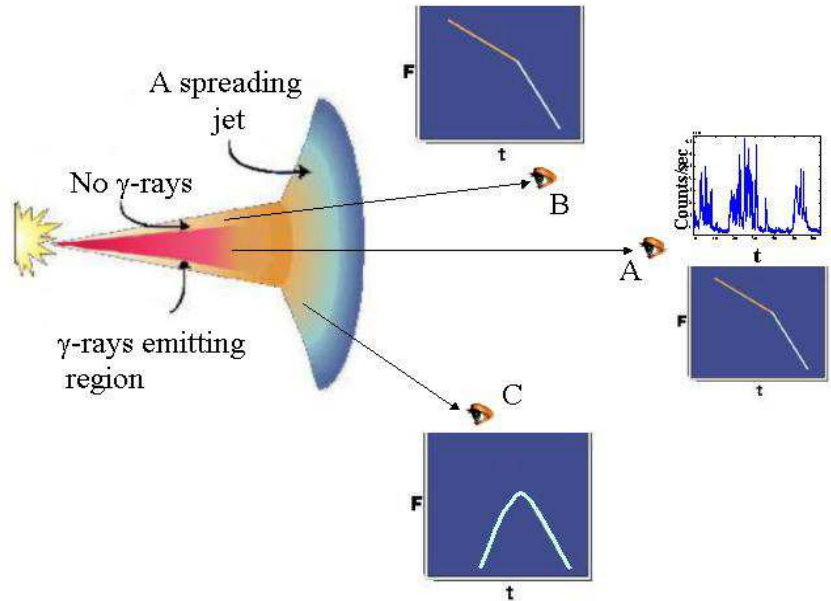}}
\par}
\caption{Off Axis Orphan afterglow is seen by observers that are
not within the initial relativistic jet. This emission is seen
only after the jet break and the sideways expansion of the
relativistic material. The schematic figure depicts three
observers. Observer A detects both the GRB and the afterglow.
Observer B does not detect the GRB but detects afterglow that is
similar to the one observed by A. Observer C detects off-axis
orphan afterglow that rises and fall and differs from the
afterglow detected by observers A and B.}
\end{figure}

A direct way to determine the beaming ratios is  to compare the
rates of detection of transients in different energy bands.
However, several confusing factors should be taken into account
in such a comparison. First, detectors in different energy bands
have different relative thresholds. These should be calibrated
using the current GRB and afterglow observations. Second, there
are numerous background transients and we have to identify
specific transients as afterglows. We show in  section 3 that
this problem may not be severe for the X-ray band. Even assuming
that all observed transients (after some basic filtering) are
afterglows we find a tight constraint on the ratio of X-ray to
$\gamma$-ray beaming. Optical background transients (e.g. AGNs,
stellar flares etc.) are more numerous. Here, we should use the
temporal and spectral observations of the afterglows, observed so
far, as templates for identification.

\begin{figure}

\label{fig:lightSchematic} {\par\centering
\resizebox*{0.9\columnwidth}{0.4\textheight}{\includegraphics{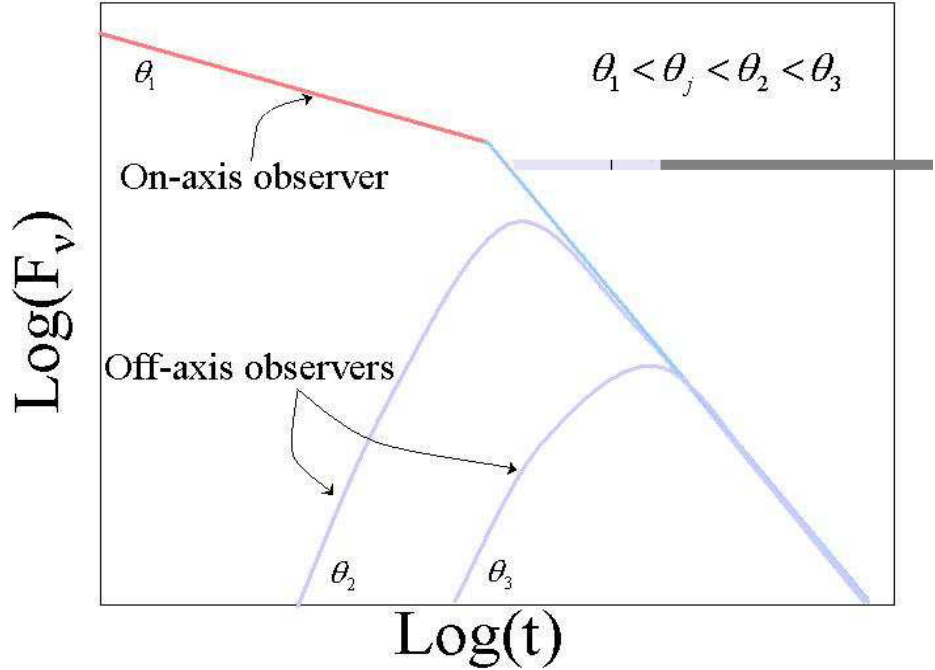}}
\par}
\caption{Schematic light curves of on-axis and off-axis orphan
afterglows. Naturally, early, when the observers is outside the
initial beam, off-axis afterglows are much weaker. Later, after
the jet break both light curves are similar.}
\end{figure}

A third problem that is unique to afterglows is the possible
confusion between the optical and radio\footnote{The current
X-ray observations are before the jet break, when only on-axis
afterglows can be seen.} on-axis and off-axis orphan afterglows.
The overall light curves of on-axis and off-axis orphan afterglows
are significantly different (see Fig \ref{fig:lightSchematic}).
However, the post-jet-break light curves of both kinds of orphan
afterglows are similar (Granot et al. 2002). To avoid confusion
we must catch the afterglow early before the jet break. This can
be done by an appropriate choice of the magnitude of an optical
survey. While off-axis afterglows are more numerous, most
off-axis orphan afterglows won't be detected in a shallow survey
and the on-axis orphan afterglows would govern such a sample. In
the radio, the transients would  be generally detected after the
jet-break and off-axis orphan afterglows would always govern the
sample. An estimate of the rate of transients in a radio survey
could not constrain the initial radio beaming, but rather it
would provide a measure of the total rate of relativistic
ejection events (Perna \& Loeb, 98; Levinson et al.,2002).

Our first goal in this paper is to estimate the X-ray to
$\gamma$-ray beaming factors using  the observed limits on X-ray
transients. We show (in section 3) that while there is a weak
indication that the $\gamma$-rays are beamed by a factor of a few
relative to the X-rays, even current observations rule out the
possibility that the $\gamma$-rays are  significantly (more than
a factor of ten) beamed than the X-rays.   There is no available
data on relevant optical transients. In a second part of the
paper (section 4) we describe possible optical surveys that, with
existing hardware, could limit the optical beaming within several
months of observation. We also discuss (in section 5) the
implications of observations of radio transients. We summarize
our results in section 6.

The observed GRB rate is a basic reference to which we compare
the rate of other transients. So far all the detected afterglows
are of long bursts (duration longer than 2sec). We exclude,
therefore, short GRBs when estimating the GRB rate. We take the
(all sky) rate of long GRBs as 600 per year (Fishman \& Meegan,
1995).

\section{Afterglow Theory}

We summarize here the relevant issues from afterglow theory.
According to the standard internal-external fireball model
internal dissipation (shocks) within ultra relativistic ($ \Gamma
>100 $) wind produces the prompt GRB. A fraction of the
prompt X-ray emission arise from this shocks as well. An
external-shock between the wind and the surrounding matter
produces the afterglow. Initially X-rays dominate the afterglow
emission, contributing to the prompt X-ray emission on time
scales of minutes and dominating the afterglow on a timescale of
hours. As the matter slows down the peak energy of the emission
shifts downwards. The X-rays are followed by optical and IR on a
time scale of days, and on a scale of weeks the dominant emission
is in the radio.

The afterglow emission is well described by a simple model of a
relativistic blast wave propagating into an external medium and
emitting via synchrotron radiation. A relativistic spherical blast
wave with energy, E, propagating in an ambient medium with a
density, n, becomes, after a short radiative phase, adiabatic
and  it follows the Blandford \& McKee (1976) self similar
solution. The Lorentz factor of the bulk motion is:
\begin{equation}
\Gamma (t)=6(\frac{4\pi
 {\varepsilon}}{10^{52}}/n)^{1/8}(t/(1+z))^{-3/8}_{day},
\end{equation}
where ${\varepsilon}=E/4\pi$ is the energy per solid angle at the
adiabatic phase, t is the observer time and z is the redshift. For
a slow cooling synchrotron radiation  the flux (above the cooling
frequency) scale with time and energy as (Sari, Piran \& Narayan
1997):
\begin{equation}
 F_{\nu }\propto {\varepsilon}^{(2+p)/4}t^{-(3p-2)/4}\; \quad \nu >\nu _{c}\, \label{Fx}
\end{equation}
where $ p\approx 2.2 $ is the electrons' energy distribution
index. After several months the blast wave becomes sub
relativistic and  then it follows the Sedov-Tylor self-similar
solution.

As noted earlier, the inhomogeneity scale in the moving ejecta
could be at most $ \Gamma^{-1} $. Moreover, an observer can see
only regions with this ($\Gamma^{-1}$) angular  size. As the
blast wave decelerates $\Gamma$ decreases, the homogeneous
regions grow  and an observer detects emission from larger
regions. Therefore, the fluctuations between observers viewing,
within the initial jet, the same burst from different angles are
expected to be largest during the $\gamma$-rays phase, decrease
in the X-rays and eventually diminish  in the optical and radio.
This is the essence of the patchy shell model (Kumar \& Piran,
00) that suggests that the large variations in the observed
$\gamma$-rays fluxes (relative to the much smaller variations in
the X-ray fluxes) arise partially from this phenomenon.

A jet is an extreme case of inhomogeneity. At the sides of the jet
the density drops sharply. As long as the jet angle, $ \theta
_{j} $, is much larger than $ \Gamma^{-1}  $, this inhomogeneity
does not affect the jet's core and it behaves as a part of a
spherical blast wave with a Lorentz factor $ \Gamma  $ and energy
$ E=4\pi {\varepsilon} $.  As $ \Gamma $ drops to $ 1/\theta
_{j}$ the jet spreads sideway (Rhoads 1997). At this stage, an
observer whose line of sight was inside of the initial jet would
see a ``jet-break" in the light curve. An observer whose line of
sight was slightly outside of the initial jet would observe at
this stage a rising flux (see figures 2 and
\ref{fig:lightSchematic}). The flux peaks when the expanding jet
gets into the observer line of sight and then it would decay
similarly to the on-axis light curve (Granot et al, 2002).

Inhomogeneity within the jet could be irregular and random, as
would be the case in a turbulent flow. Small scale inhomogeneities
within the jet (e.g. hot spots) could cause  a modulation of the
observed $\gamma$-ray flux relative to the X-rays and the optical
(see fig \ref{fig:OnAxisOrph}). As the jet slows down the causally
connected regions, that are of order $\Gamma^{-1}$, increase and
the jet becomes smoother.   At the same time as the jet slows down
the typical emission energy decreases. This will produce
different beaming factors for the different energy bands with
wider beams at lower energies.

Alternatively, Rossi, Lazzati and Rees (2002) consider a smoothly
varying jet with  ${\varepsilon} ={\varepsilon} (\theta )$, and
$\Gamma = \Gamma (\theta )$. As long as $ \Gamma (\theta _{1})\gg
1/\theta _{1} $ the behavior of a part in the jet at an angle $
\theta _{1} $ is similar to a the behavior of a sphere with a
constant $ {\varepsilon} ={\varepsilon} (\theta _{1})$. When
$\Gamma (\theta_1 ) \approx 1/\theta_1$, the jet at $ \theta _{1}
$ spreads outwards. Rossi et. al (2002) show that before the
spreading, the emission observed  at an angle $ \theta _{1} $ is
dominated by the part of the jet that is along the line of sight.
Therefore, before the sideways spreading an observer at
$\theta_1$  will see a similar light curve to an observer of a
homogeneous jet with $ {\varepsilon} ={\varepsilon}(\theta
_{1})$. The beaming at different wavelengths could be different.
As the physical conditions vary  with $\theta$ it is possible
that an emission at some energy bands won't be produced
throughout the whole jet. For example, if the Lorentz factor at
certain angles is too low the region may not emit $ \gamma $-rays
but would still produce X-ray and optical emission (Rossi et. al.
2002).

\section{X-ray Transients}
\label{sec:xrays}

The GRB X-ray emission is divided to two: prompt and afterglow (or
even three: prompt, early afterglow and late afterglow)
components. Late X-ray afterglow emission was  detected from
almost all BeppoSAX bursts  during the first hours of the
afterglow with:
\begin{equation}
F_{\nu}=6^{+4}_{-2} \times 10^{-13} ({\frac {t}{11{\rm
hr}}})^{-\alpha} {\rm ergs/sec/cm}^2 , \label{xrays}
\end{equation}
where $\alpha \sim 1.3 \pm 0.3$.   The X-ray fluxes 11 hours
after the burst are narrowly clustered (within a factor of $\sim
7$) around this average value (Piran et. al. 01). Prompt X-ray
emission at the level of $10^{-8}$ergs/sec/cm$^2$ accompanied the
GRBs themselves. The prompt X-ray emission can be divided to two
parts: the low energy tail of the $\gamma$-ray signal and a
softer and smoother later component that looks like the beginning
of the afterglow phase (Piro, 00). Within the internal-external
shocks model these two components can be interpreted as emission
from internal and external shocks respectively.

Before considering  X-ray surveys we examine some general
arguments concerning the detection rate and the implied event
rate.  Three time scales arise when the number of detected
bursters in a survey is interpreted as a burster rate: The
interval between successive observations of the same field,
$\Delta t$, the duration of each observation, $\delta t$, and the
time  that the transient is above the survey's threshold,
$t_{obs}$. Consider a survey with a sensitivity, $F_{lim}$,  that
covers a field of view, $A$, for $N_s$ times in intervals of
$\Delta t$. The burster flux, $F(t)$, decays with time with
$F(t_{max})=F_{lim}$. The expected number of events in the survey
is:
\begin{equation} N=R A N_s {\rm max}[ \delta t, {\rm min}(\Delta
t, t_{max})] ,
\end{equation}
where  $R$ is the true burster's rate (events per time unit per
solid angle).

Greiner et. al. (1999) searched the ROSAT all sky survey (RASS)
for afterglow candidates at a limiting flux of $
10^{-12}$ergs/sec/cm$^{2}$. This limit corresponds to the
afterglow flux observed several hours after the burst when,
according to  the standard afterglow model, $ \Gamma=10-20 $. The
sky coverage of RASS is 76435 square degree $\times$ days, and
80\% of the afterglows in the field of view are expected to be
detected. The observed  long GRB rate,  600 per year on the whole
sky, implies 3 expected events in the RASS field (assuming only
long GRBs have afterglow). Greiner et. al. (1999) find at most 23
GRB afterglow candidates. This suggests at first  a ratio of 8 in
the beaming factors. However, an examination of the optical
emission of six randomly chosen transients (from this sample of
23) revealed that all six are flaring stars. Greiner et. al.
(1999) conclude that the majority of the 23 sources are not GRB
afterglows. This suggests a moderate difference (if any) in the
beaming between late (several hours) X-ray afterglow and
$\gamma$-ray emission.

Grindlay (1999) analyzed the results of the ARIEL 5 X-ray
transients. The survey's sensitivity is $ 4\times 10^{-10}$
ergs/sec/cm$^{2}$ and the data is collected at time bins of
100min ($\Delta t=\delta t=100$min). The extrapolation of
BeppoSAX's observations (eq. \ref{xrays}) suggests that the flux
level of  $ 4\times 10^{-10}{\rm ergs/sec/cm}^{2} $ is reached $
\sim 10$min after the burst. Therefore it is unlikely that an
afterglow was detected in more than one time bin.  Moreover,
since each time bin is the sum of the emission during a period
of  100min, the triggered time bin includes the prompt emission
as well as the comparable or weaker initial afterglow. Therefore,
the transients observed in this survey are governed by the prompt
emission. Grindlay (1999) finds 13 candidates  (9 of them
detected only in a single time bin). This corresponds to a rate
of 0.15 transients per day over the whole sky. This rate is
consistent with the logN-logS relation in the BATSE catalog,
assuming that the fluence in $ \gamma $-rays is 5 times the
fluence in X-rays (note that this kind of association with
BATSE's logN-logS is relevant only if the observed emission is
governed by the prompt emission). This result suggests no
difference in the beaming of the prompt X-ray and $ \gamma $-ray
emission. However it does not constrain the beaming of the late
(lower $\Gamma$) X-ray emission.

\begin{table}
{\centering \begin{tabular}{|c|c|c|c|c|c|c|c|} \hline
Survey&Sensitivity& \( \Delta t \)& \( \delta t \)& \( t_{max} \)&
Expected & Observed & Beaming\\
&\( {\rm erg/sec/cm}^{2} \)&  min &  min & min& counts&
counts &  ratio $ \ ^{(***)}$\\
\hline \hline ROSAT& \( 10^{-12} \)& \( 96 \)& \( 0.16-0.5 \)&
400& 3&$<$17 &  $\ll 8$\\
\hline HEAO-1& \( 7\times 10^{-11} \)& \( 35 \)& \(
0.16 \)& 20& 2.5&$<$4 & $<3$\\
\hline Ariel 5& \( ^{(*)}~4\times 10^{-10} \)&
\( 100 \)& \( 100 \)& prompt & 10-20&$<$13 & $<1$ \\
\hline BeppoSax WFC& \( 10^{-8} \)& -& -&prompt& $\ ^{(**)}~49$&
$<$66 &  $ < 1.34\pm .23$\\
\hline
\end{tabular}\par}

\caption{ Results from X-ray transients surveys. For each
  survey  we list the sensitivity, the interval between following
observations of the same field, $\Delta t$ ; the duration of each
observation, $\delta t$, and the averaged time after the burst
that an X-ray afterglow is above the survey's threshold,
$t_{max}$. The expected counts are calculated from the long GRB
rate. Note that the different surveys are complementary as they
explore different phases (different sensitivity and therefore
different $t_{max}$ values) of the X-ray afterglow.}

(*)-This flux is averaged over 100min. The minimal total observed
 energy in 100min needed for detection is $2.4\times 10^{-6}$ergs/cm$^{2}$.

(**)-The expected rate is given according to the observed number
 of GRBs by the WFCs. The observed rate includes the XRFs.

(***) These upper limits  on the ratio of X-ray to $\gamma$-ray
beaming are at 5\% confidence level. The limits were obtained
assuming that all observed unidentified transients are related to
GRBs. This last assumption is clearly a gross over estimate for
the RASS survey.

\end{table}

Grindlay (1999) also analyzed the results of HEAO-1 survey
(Ambruster \& Wood, 1986). HEAO-1 observe each source for $\delta
t=10$sec in time intervals of $\Delta t = 35$min. The sensitivity
of the survey, $ 7\times 10^{-11}{\rm ergs/sec/cm}^{2}$,
corresponds to the observed X-ray fluxes $ \sim 20$min after the
burst. Since  the duration of the prompt emission  is $\sim 1$min
it is more likely that an afterglow emission would be detected.
With $t_{obs}=20$min and $\Delta t=35$min only 1 out of any 1.75
afterglows would be detected. Therefore, we interpret the four
unidentified (out of 10) detected transients as afterglow
candidates (the other 6 have another clear identification).
Ambruster \& Wood (1986) deduce from the 10 observed transients,
assuming that the duration of the transients $>35$min, a rate of
1500yr$^{-1}$ transients in the whole sky. This rate corresponds
to $1500 \times (4/10) \times 1.75 = 1050$yr$^{-1}$ X-ray
afterglow transients over the whole sky. This rate suggests again
a very modest difference between the X-ray and $\gamma$-ray
beaming. One of the 4 candidates in the HEAO-1 survey shows
variability on the 10sec observation time scale, a very hard
spectrum and a flux of $4 \times 10^{-9}$ergs/sec/cm$^2$. This
suggests that the transient is probably a prompt emission. This
interpretation suggests  a large event rate. As a prompt emission
takes 1min ($\Delta t / t_{max} \sim 30$) the expected rate for a
single detection is $1500\times (1/10) \times 30 = 4500
$yr$^{-1}$ over the whole sky, seven and a half times the long
GRB rate and five times the overall GRB rate.  However,  the
identification as prompt emission is uncertain and the
statistical fluctuation with a single event could be large.

BeppoSAX observed the early (first $\sim 100$sec) X-ray emission
with the WFCs and the late ($ \sim  $1day) X-ray afterglow with
the NFI. Almost in all cases when a prompt emission was observed
in the GRBM, an X-ray afterglow was detected by the NFI as well
(Piro, 00). The WFC  is capable of detecting events of a few
seconds at a level, $\sim 10^{-8}$ergs/sec/cm$^2$, comparable to
the prompt X-ray emission accompanying the GRB. For integration
times of ~100 sec, transients can be detected down to $\sim
10^{-9}$ergs/sec/cm$^2$,  allowing to track the the early
afterglow for a few hundreds seconds. The WFCs detected X-ray
counterparts of GRBs in 49 cases when the GRBM was triggered. In
39 cases the WFCs were triggered without GRBM triggering. The
duration of 17 of these transients (out of the 39 transients),
denoted X-ray flashes (XRFs), is comparable to the duration of
the X-ray emission accompanying GRBs  (Heise et al, 01).   The
peak fluxes of the XRFs are similar to the X-ray fluxes observed
during GRBs in the WFCs ($\sim 10^{-8}$ergs/sec/cm$^2$). The
origin of XRFs is not clear.

In the context of this paper, we can adopt the working hypothesis
that  XRFs are prompt X-ray emission with no accompanying
$\gamma$-ray emission (namely, they are associated with GRBs that
have been pointing elsewhere). This assumption is not necessarily
true. XRFs may be the early afterglow emission (but this is
unlikely because they are rather variable while this early
afterglow is expected to be smooth). Alternatively XRFs may be an
unrelated phenomenon. Our working hypothesis results in an upper
limit to the ratio of prompt X-ray emission to GRBs.  If this
interpretation of XRFs (or a fraction of them) is correct the
prompt X-ray rate is comparable to the observed GRB rate.  If we
identify all XRFs as associated with GRBs we find that the ratio
of beaming of X-ray to $\gamma$-rays is (at 5\% confidence level)
$1.34\pm .23 $. If this association is incorrect than this ratio
is just an upper limit. Overall, the WFCs' observations imply that
the beaming factors of the prompt $ \gamma $-rays and the prompt
early X-rays  are comparable, up to a factor of two.

After a GRB localization  BeppoSAX redirected its NFI to observe
the late afterglow. There is a several (at least three) hours gap
(the redirecting time) between the two observations. The NFI
afterglow observations last for about a day, until the flux drops
below the NFI's sensitivity. The light curve during the X-ray
afterglow is fitted well with  a single power law. Remarkably the
extrapolation of this power law to the early afterglow's results
in fluxes  similar to those observed at that time by the WFCs
(Piro, 00). According to the standard afterglow model $ F_{x } $
is almost linear with $ {\varepsilon} $ (see eq. \ref{Fx}). The
single power law decay of $F_x$ through the whole evolution
implies that the effective $ {\varepsilon} $  is constant, or at
least it varies regularly (as a power law in t) while the
observed area varies by two orders of magnitude, from
(0.01rad)$^2$ after a minute to (0.1rad)$^2$ after a day. This
result suggests a regular structure and possibly a similar
beaming of the late and early X-ray afterglows.

To summerize the X-ray results, BeppoSax and Ariel 5  provide a
tight constrain on the prompt and early afterglow X-rays to
$\gamma$-rays beaming ratio (not more than a factor of two). The
HEAO 1 results constrain the beaming ratio of the intermediate
(about 20min after the burst and a corresponding Lorentz factor
of ~30) afterglow to $\gamma$-rays  to be smaller than 3.
Finally, the RASS results constrain the beaming ratio of  the late
(several hours and a corresponding Lorentz factor of ~10) X-ray
afterglow to $\gamma$-rays to be smaller than 8 and probably much
less. All these limits are relevant to on-axis orphan afterglows
as at this early stage the X-ray fluxes from off-axis orphan
afterglows would be much lower  and could not be detected (see fig
\ref{fig:lightSchematic}).

\section{Optical Transients}

There are no current deep searches that have systematically looked
for (either on-axis or off-axis) optical orphan afterglow
transients (see however, Vanden Berk et al., 2002
 for a ongoing search using the SDSS).
Note, however, that some useful upper limits might be obtained
from Supernovae searches and from Near Earth Objects (NEO)
searches. We describe, in this section, the characteristics of
optical on-axis orphan afterglows and show that an orphan
afterglow survey is feasible with current equipment.

Several problem arise when we turn to compare the rate of optical
afterglow transients to the rate of GRBs.  First there are
numerous other transients that should be excluded. Second, not
all well localized GRBs have a detectable optical counterpart
(optical afterglow was observed only in $\sim$50\%  of the cases
with good localization). Third, there may be confusion between on
axis and off axis orphan afteglows. These problems suggest that it
will be impossible to obtain an exact estimate of the optical
orphan afterglow rate. But as we have seen in the X-ray case, for
our purpose an upper limit on the rate could also be very useful.

\begin{figure}
\label{fig:MagOneDay} {\par\centering
\resizebox*{0.9\columnwidth}{0.3\textheight}{\includegraphics{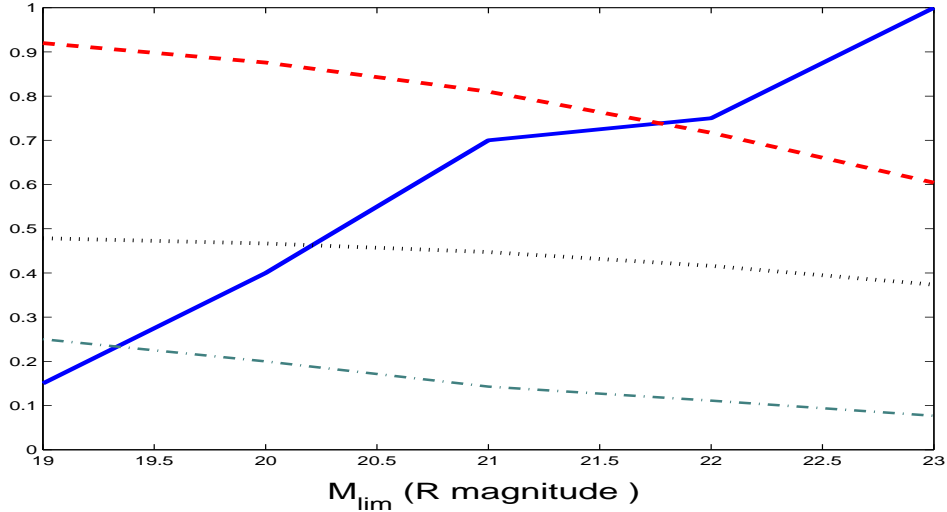}}
\par}
\caption{The cumulative fraction of optical afterglows brighter
than  R magnitude   $m_{lim}$ one day after the burst ({\it solid
  line}). The optical afterglows sample includes  20 afterglows with good
data. The ratio of on-axis to off-axis optical afterglows are
according to Nakar Piran \& Granot (2002) model A ({\it dashed
line}) and model B ({\it dotted line}), and according to Totani \&
Panaitescu (2002) ({\it dashed-dotted line}). }
\end{figure}

We ask what should be the depth and the covering area of an ``on
axis" afterglow survey. Clearly it should not be too shallow so
most of the on-axis afterglows will be detected. It shouldn't be
unnecessarily deep as this would increase the rate of other
background transients and will also include a large number  of
off-axis afterglows that will dominate the sample. The jet break
time in most observed afterglows is $ \sim 1 $day. At this stage
the observed R magnitude is about 19th-23rd. Figure
\ref{fig:MagOneDay} depicts the cumulative fraction of afterglows
(out of the observed ones) with R magnitude brighter than a
limiting magnitude after one day. At this time only 15\% of the
observed afterglows, are brighter than R=19th mag while 70\% of
the observed afterglows are brighter than 21st mag. 40\% of the
afterglows were brighter than 21st mag after two days, enabling
dual detection in a repeating survey night after night.

Figure \ref{fig:MagOneDay}  also depicts the expected ratio of
on-axis afterglows to off-axis afterglows  as a function of the
observed magnitude (Nakar, Piran \& Granot 2002; Totani \&
Panaitescu 2002). While different models give different ratios of
off-axis to on-axis orphan afterglows, all show that at  a
19th-21st limiting magnitude the rate of off-axis afterglows is
at most comparable to the on-axis afterglow rate and it is most
likely that the on-axis afterglows are the majority. Therefore,
we conclude that the survey should have a minimal limiting
magnitude of 19th. A preferred limit should be 21st, but not
deeper.

\begin{figure}
\label{fig:M-t} {\par\centering
\resizebox*{0.9\columnwidth}{0.3\textheight}{\includegraphics{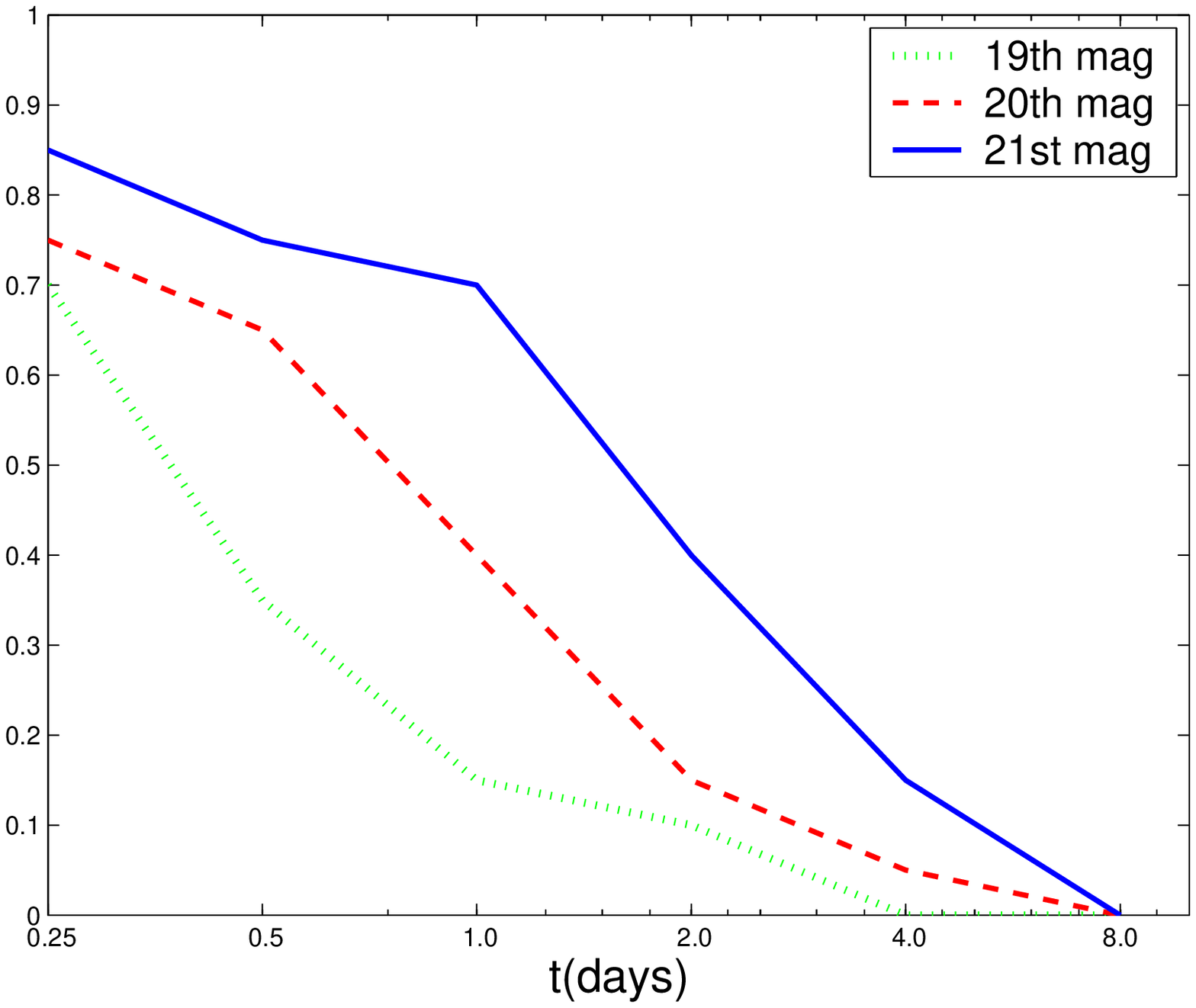}}
\par} {The cumulative fraction of optical afterglows with R
  magnitude greater than 21st ({\it solid
  line}), 20th ({\it dashed line}) and 19th ({\it dotted line}),
magnitude as a function of time. The optical afterglows sample are the
same  afterglows as in figure \ref{fig:MagOneDay}}
\end{figure}

We turn now to estimate the covering factor (area $\times$ time)
needed for  meaningful results.  This estimate can be discussed
in terms of the general survey discussed in section
\ref{sec:xrays} with  $\Delta t= \infty$, $\delta t = 0$ and with
$t_{max}$ averaged over the current optical afterglow sample.
Assuming that only half of the long bursts are followed by an
observed optical counterpart, there is one optical afterglow
associated with $\gamma$-ray emission every day. A 19th mag
snapshot is expected to record $10^{-5}$ afterglow transients
associated with GRBs per square degree. This is a weighted
average over the current observed afterglows that takes into
account that each burst contributes to the detection probability
according to the time it is above 19th R magnitude. Although only
15\% of the bursts have 19th R magnitude a day after the burst,
the average weighted detection fraction is 40\%.  A   21st mag
snapshot is expects to record  $4 \times 10^{-5}$ afterglow
transients per square degree. Figure \ref{fig:M-t} depicts the
fraction of afterglows with flux brighter than 19th, 20th and
21st mag as a function of time after the burst. We  expect $2
\times 10^{-5}$ dual records of the same afterglow per square
degree in two consecutive 21st mag snapshots of the same field 24
hours apart. Any detection of more than a factor of a few above
these expected rates would indicate a different beaming of the
optical emission and would have meaningful implications to our
understanding of GRBs.

To avoid contamination by other transients the survey should
include a follow up procedure that will establish that a given
transient is a GRB afterglow. A critical  decision is which
transients should be  followed.  The number of transients at
these magnitudes is unknown (e.g. Paczynski, 01), but they are
probably too numerous to follow. A significant group of
misleading sources could be flaring sources (stars or AGNs) with
a steady flux lower than the survey sensitivity. Such sources
could be discarded by performing  the survey on a field observed
by the SDSS (York et al., 2000) that shows all the steady sources
up to 23rd R magnitude. Alternatively one can cover once the
observed region with a deeper exposure, that could serve as a
reference. Unless the flare is extremely bright the steady source
would be seen  in the deeper survey (e.g. SDSS)  and could be
automatically eliminated. The light curve templates of existing
afterglows would provide another cut. 40\% of the observed
afterglows are brighter than 21st  R magnitude two days after the
burst. Therefore, a repeated observation of the same region after
24 hours  should show the decaying source again with a lower flux
(typically 1-2 magnitudes).   This method could not be used in a
19th mag survey, for which only a single detection is expected
for most afterglows. A repeated long survey of the same region
would also be advantageous as we could learn the history of this
field and eliminate repeated transients.

The SDSS (York, et al., 2002) is too deep and covers too small
area for on-axis orphan afterglow search (see Nakar et al.,
2002). An automated wide field survey could cover a field of view
equal to the whole sky in several months. This should result in
one to two optical afterglow transients associated with GRBs. The
Super-LOTIS \break
\url{http://compton.as.arizona.edu/LOTIS/super.html}, for example,
covers in one minute of exposure 0.7 square degree in 21st
magnitude. Therefore a few months of continuous Super LOTIS
observation could constrain the beaming difference up to one
order of magnitude. NEOs searches scan the whole sky at $\sim
20$th magnitudes. The LONEOS \break
\url{http://asteroid.lowell.edu/asteroid/loneos/loneos1.html}
covers 20000 square degrees in one  month at V=19.3. In 5 months
of observations one afterglow associated with GRB should appear in
the records.  An order of magnitude constraint could be obtained
within several weeks. Note that the filtering of transients in
this survey might be more difficult as this survey does not
observe the same field night after night, morever  it does not
observed the field  already covered by the SDSS. Both surveys
would needs a dedicated follow up telescope that will observe the
spectrum and will follow the light curves of the transients in
order to confirm a detection of a transient as an afterglow.

The proposed LSST, Large-aperture Synoptic Survey Telescope (Tyson,
Wittman \& Angel 2000), is an 8.4m telescope with a 7 square
degrees field of view. It can cover 20000 square degrees during
3-4 clear nights to a limiting magnitude of V=24. This telescope
could  find all transients with a 21st limiting magnitude and
follow their light curves for more than a week. Its deep field
will enable an easy discard of wrong candidates. It can also
obtain an 8 color image that  will help identifying transients as
afterglows. A single week of observations with the LSST would
give a tight constrain on the optical beaming. Another exciting
result possible with this telescope is the comparison of the
observed afterglows at 21st and 24th magnitudes. This will enable
a determination of the ratio of on-axis to off-axis afterglows
and will provide another handle on the ``optical beaming factor".

\section{Radio Transients}

The radio afterglow flux is still rising a few days after the
burst. Therefore, in any survey the dominant population of radio
afterglows would be from off-axis orphan afterglows. Hence, there
is no way to constrain the initial  radio beaming by the observed
rate of radio afterglows. However, as suggested by Perna \& Loeb
(1998), the rate of the off-axis orphan radio afterglows can
determine the total rate of GRBs or rather the rate of
relativistic ejecta events, regardless of the prompt emission in
different wave lengths at different stages.

After about 100 days the jet becomes spherical and
sub-relativistic (Waxman, Frail \& Kulkarni 1998). At this stage
the only relevant initial condition is the total energy, $E_K$, of
the jet at the beginning of the adiabatic phase. This determines
the strength of the emission and hence the distance from which the
late radio afterglow can be detected.  The rate of detection of
radio afterglows  depends, therefore,  on the initial energy and
on the rate of GRBs. However, both parameters (total energy and
rate) are unknowns. Levinson et. al. (2002) estimate $E_{K}$ as
$E_{iso} f_b$ and the true GRB rate as $ f_b^{-1}$ times the
observed GRB rate. They find that the expected detection rate of
radio afterglows depends almost linearly on the beaming factor.
With a lower $f_b^{-1}$ there are fewer but more energetic events.
Those more energetic events can be detected from larger distances
(typically of order of a few hundred Mpc). The gain in the
detection volume overcomes the loss due to the lower rate. Using
the FIRST and NVSS surveys Levinson et. al. (02) find at least 25
orphan afterglow candidates. This set a limit of $f_{b}^{-1}>10
$. If none of these transients are orphan afterglows than beaming
will be stronger with $f_b^{-1} > 100$.

However,  in the context of our paper,  we allow for different
beaming factors in different energy bands and the above relations
are not trivial. There may not be a single parameter, $f_b$, that
determines both the rate of afterglows and their energy. With
multiple beaming factors, the $\gamma$-rays beaming factor,
$f_{b_{\gamma}}$, would determine the expected rate of orphan
radio afterglow relative to the rate of GRBs. However, the total
energy during the mildly relativistic radio phase may be much
larger than $E_{\gamma iso} f_{b_\gamma}$. A large amount of
energy could be `hidden' in a relativistic component that does not
emit $\gamma$-rays or even X-rays. This may be a problem, as the
expected rate  depends on the 11/6 power of the total energy
(Levinson et. al. 2002). Thus one can use the Levinson et al.
(2002) limits only if there is an independent evidence that
$f_{b_\gamma} \sim f_{b_R}$. A determination that $f_{b_\gamma}
\approx f_{b_{opt}}$ would be a step in the right direction.

\section{Conclusions}

There are various reasons to expect that GRB jets will have an
angle dependent profile. For example, this is  natural in the
Collapsar model, in which a jet punches a hole in the surrounding
stellar envelope. One can expect that along with the
ultra-relativistic motion at the core of the jet there would be
slower motion of a thicker envelope. The relations between the
different beaming factors have many implications on different
parameters of the GRB, most notably on the total energy budget.
For example Frail et. al. (2001) and Panaitescu \& Kumar (2001)
calculate the total energy emitted in the prompt emission using:
$ E_{\gamma }=E_{\gamma iso}f_{b_{opt}} $, where $ E_{\gamma iso}
$ is the isotropic energy in $ \gamma $-rays (during the several
dozen seconds of the burst) and $ f_{b_{opt}}=\theta^2_j/2 $ is
the beaming factor of the optical emission several days after the
burst. Clearly, this calculation is relevant only if $
f_{b_{opt}}\approx f_{b_{\gamma }} $, namely if the beaming
during the afterglow after a day or so reflect the beaming during
the prompt $\gamma$-ray emission. Indeed this fact that $
E_{\gamma }=E_{\gamma iso}f_{b_{opt}} $, is narrowly distributed
suggests that $f_{b_{opt}}\approx f_{b_{\gamma }} $ (or at least
$f_{b_{opt}}\propto f_{b_{\gamma }} $). Otherwise this narrow
distribution would require a miraculous coincidence.

 These ideas have motivated our
study of the possibility of different beaming factors for
different observed energies that correspond to matter moving with
different Lorentz factors. Our analysis shows that the results of
the three  X-ray surveys, as well as the BeppoSAX data are
consistent with a very moderate difference in the $\gamma$-rays
and X-rays beaming. The four surveys shows that both the early
and the late X-ray beaming, $f_{b_x}$, are comparable to the
$\gamma$-ray beaming, $f_{b_\gamma}$. Our most stringent limit
$f_{b_x} < (1.34 \pm .23)  f_{b_\gamma}$, is obtained  for the
prompt X-ray emission from the BeppoSAX WFCs. This is only an
upper limit because of the uncertainty in the interpretation of
the XRFs as orphan GRB counterparts. The ratio is even lower if
XRFs are not related to GRBs.

The highest limit is obtained for the afterglow 400 minutes after
the burst from the ROSAT RASS data. With 17 unidentified
transients compared to 3 expected we find $f_{b_x} < 8
f_{b_\gamma}$. However, this is only a weak upper limit as all 6
transients  (out of the original 23 transients) that have been
checked  were found to be flaring stars. It is possible that all
the remaining 17 transients are not related to GRBs. We conclude
that during the X-ray afterglow (20-400 minutes after the burst)
$f_{b_x}$ is probably less than twice $f_{b_\gamma}$ and $f_{b_x}
\approx 10 f_{b_\gamma}$ is certainly ruled out by current
observations. This result shows that the bulk energies at
$\Gamma=200$ and at $\Gamma =10$ are comparable. This result
supports the homogeneous jet approximation. It puts a strong
constraint on the $\theta$ dependent jet model of Rossi et al.,
(02), as it requires that the ratio of $\gamma$-ray to X-ray
emission is roughly constant throughout this variable jet.

In our analysis we have implicitly assumed (by using Eq.
\ref{xrays}) that the energy per unit solid angle is similar in
the on-axis orphan afterglows to the one in regular GRB
afterglows. If this energy is lower by a given factor
$\varepsilon_x/\varepsilon$, then using Eq. \ref{Fx} we find that
$t_{max}$ will decrease by a similar factor. The signal will be
detected but for a shorter period, as long as the maximal X-ray
flux is above the sensitivity limit of the survey. This, in turn,
will decrease the detection rate by $\varepsilon_x/\varepsilon$
and will increase the implied limits on the ratio of $\gamma$-ray
beaming to X-ray beaming by $(\varepsilon_x/\varepsilon)^{-1}$.
The total energy in X-ray emitting matter is proportional to
$f_{bx} \varepsilon_x$. Hence our results impose a direct limit
(independent of $\varepsilon_x$) on this quantity. They show that
the total energy in X-ray cannot be significantly larger than the
total energy in $\gamma$-ray emitting matter.

It will be a remarkable achievement to constrain also the optical
($\Gamma \sim 2-5$) beaming. We have argued that even existing
hardware like Super-LOTIS can rule out (or confirm)
$f_{b_{opt}}\approx 10 f_{b_\gamma}$ on a time scale of several
months. The proposed LTSS can constrain this factor within a week.
Radio observations provide a different limit on the total rate of
relativistic ejecta events. However, this limit can be obtained
only under the assumption of a single beaming factor. If optical
afterglow observations would show that $f_{b_{opt}} \approx
f_{b_{\gamma}}$ this assumption would be reasonable. Under this
assumption the comparison of the radio orphan afterglow rate to
the GRB rate can set limits on $f_b$. Levinson et al (02) find
that current surveys limit $f^{-1}_b > 10$. This is consistent
with the rest of the observational results.

One argument that supports a beaming ratio of a few is obtained by
comparing the different energies during the GRB. A detailed
analysis of the afterglow light curves (Panaiteschu \& Kumar, 01)
and of the X-ray fluxes (Piran et al., 01)  have shown that $E_K$
the kinetic energy during the adiabatic afterglow phase is also
narrowly distributed. Panaiteschu \& Kumar (01) find  using $
E_{\gamma }=E_{\gamma iso}f_{b_{opt}} $ and their estimates of
$f_{b_{opt}}$ that the average $E_\gamma$ is larger than the
average $E_K$: $\bar E_\gamma \approx 3 \bar E_K$. This result is
somewhat puzzling (Piran, 01): A narrow $E_\gamma$ distribution
with $\bar E_\gamma
> \bar E_K$ implies (with no fine tuning) that  $E_{rel}$ (the
total energy of relativistic ejecta) is narrowly distributed and
$E_\gamma \approx E_{rel}$ (rather than $E_{rel} \approx E_K$).
The fact that $E_K<E_\gamma$ is also narrowly distributed implies
that $\epsilon_{\gamma}$, the conversion efficiency of
relativistic kinetic energy to $\gamma$-rays ($E_{\gamma} \equiv
\epsilon_{\gamma} E_{rel}$), is close to unity and moreover,
$\epsilon_{\gamma}$ itself should be rather narrowly distributed
(between 70-80\%). This last conclusion would be astonishing
considering the dependence of $\epsilon_{\gamma}$ on the the
distribution of energies and Lorentz factors of the different
shells that produce the internal shocks. This puzzle can be
resolved if according to the patchy shell model, the $\gamma$-ray
emitting regions are indeed narrower than the late time afterglow
emitting regions. In this case the quantity: $E_{\gamma
iso}f_{b_{opt}} $ is an overestimate of $E_{\gamma}$ by the
relative beaming factor. Specifically, $E_{\gamma}$ would have
been overestimated by a factor of 3 if  $f_\gamma \approx
f_{opt}/3$ (Piran, 01).

Before concluding we remark that our results also impose limits
on the rate of  ``failed GRBs".  We define here ``failed GRBs" as
events in which the source fails to produce GRB but ejects a
relativistic ejecta (comparable in energy to GRBs' ejecta) that
produces a detectable afterglow. From the point of view of this
paper ``failed GRBs" can be viewed as extreme cases of on-axis
orphan afterglows in which the $\gamma$-ray beam vanishes. In
these events our ``on axis orphan afterglows" would be true
orphans having no GRB parent pointing towards any observer. The
implied rates that we have found provide immediately limits on
the ratio of ``failed GRBs" to GRBs. In particular the ROSAT
survey shows that the rate of ``failed GRBs" with a regular X-ray
afterglow cannot be more than a factor of eight (and possibly
much lower) than the rate of GRBs. If the energy per solid angle
is smaller, then as discussed earlier, the limit on the rate would
be higher by the inverse factor.  Optical survey, of the type we
discuss here, could limit the rate of ``failed GRBs" with an
optical afterglow. This would presumably correspond to events in
which the matter is ejected with a lower Lorentz factor.

We thank Avishay Gal-Yam, Jonathan Granot, Jonathan Katz, Pawan
Kumar and Luigi Piro for helpful discussions. This research was
supported by the US-Israel BSF.

\end{document}